\documentclass[aps,prl,showpacs,floatfix,twocolumn]{revtex4}

\usepackage{graphicx}
\usepackage{amssymb,amsmath,amsfonts,amsbsy}

\newcommand{\beq}{\begin{equation}}
\newcommand{\eeq}{\end{equation}}
\newcommand{\bdis}{\begin{displaymath}}
\newcommand{\edis}{\end{displaymath}}
\newcommand{\bea}{\begin{eqnarray}}
\newcommand{\eea}{\end{eqnarray}}
\newcommand{\barr}{\begin{array}}
\newcommand{\earr}{\end{array}}
\newcommand{\bfig}{\begin{figure}[!]}
\newcommand{\efig}{\end{figure}}

\newcommand {\Npar}{N_{z}}
\newcommand {\Nperp}{N_{\bot}}

\newcommand {\lEX}{l_{_\mathit{\! EX}}}

\newcommand {\rvec}{\mathbf{r}}

\newcommand {\hvecM}{\mathbf{h}_{_\mathit{\! M}}}

\newcommand {\haz}{\text{h}_{az}}

\newcommand {\hM}{\text{h}_{_\mathit{\! M}}}

\newcommand {\Msat}{\text{M}_{s}}

\newcommand {\mvec}{\mathbf{m}}

\newcommand {\mveco}{\mathbf{m}_{0}}
\newcommand {\mvecopar}{\mathbf{m}_{0 z}}
\newcommand {\mvecoperp}{\mathbf{m}_{0 \bot}}

\newcommand {\evec}{\mathbf{e}}

\newcommand {\evecz}{\mathbf{e}_{z}}

\newcommand {\evecone}{\mathbf{e}_{1}}
\newcommand {\evectwo}{\mathbf{e}_{2}}

\newcommand {\avec}{\mathbf{{a}}}

\newcommand {\qvec}{{\mathbf{q}}}

\begin{document}

\title{Spin-wave instabilities in spin-transfer-driven magnetization dynamics}

\author{G. Bertotti$^{(a)}$, R. Bonin$^{(b)}$, M. d'Aquino$^{(c)}$, C. Serpico$^{(d)}$, I. D. Mayergoyz$^{(e)}$}

\affiliation{
$^{(a)}$INRIM - Istituto Nazionale di Ricerca Metrologica, 10135 Torino, Italy \\
$^{(b)}$Politecnico di Torino, sede di Verr\`es, 11029 Aosta, Italy \\
$^{(c)}$Dip. Tecnologie, Universit\`a "Parthenope", 80143 Napoli, Italy \\
$^{(d)}$Dip. Ing. Elettr., Universit\`a "Federico II", 80125 Napoli, Italy \\
$^{(e)}$ECE Dept., UMIACS, AppEl Center, University of Maryland, College Park MD 20742, USA}

\date{\today}

\begin{abstract}
We study the stability of magnetization precessions induced in spin-transfer devices by the injection of spin-polarized electric currents. Instability conditions are derived by introducing a generalized, far-from-equilibrium interpretation of spin-waves. It is shown that instabilities are generated by distinct groups of magnetostatically coupled spin-waves. Stability diagrams are constructed as a function of external magnetic field and injected spin-polarized current. These diagrams show that applying larger fields and currents has a stabilizing effect on magnetization precessions. Analytical results are compared with numerical simulations of spin-transfer-driven magnetization dynamics.
\end{abstract}


\maketitle

Currents of spin-polarized electrons can induce large-amplitude magnetization precessions at microwave frequencies in small-enough magnetic devices \cite{Slonczewski1996,Berger1996}. There is mounting experimental evidence that these so-called spin-transfer phenomena do occur in nano-pillar or nano-contact devices under current densities of the order of $10^6 - 10^8$ A/cm$^2$ \cite{Kiselev2003,Rippard2004,Krivorotov2007,Boone2009}. This discovery has boosted the already widespread interest in the physics of the interplay between magnetism and electron transport, and has triggered efforts toward the promising development of new generations of microwave spin-transfer nano-oscillators.

A spin-transfer device is a non-linear open system, driven far-from-equilibrium by the action of the spin-polarized electric current. The excited magnetization precessions represent strong excitations of the magnetic medium, which in principle may give rise to various types of instability and eventually to transitions to chaotic dynamics. A parallel can be drawn with ferromagnetic-resonance Suhl's instabilities \cite{Suhl1957}, in which certain spin-waves can get coupled to the uniform precession and start to grow to large non-thermal amplitudes, thus destroying the spatial uniformity of the original state.

In this Letter, we demonstrate that spin-wave instabilities may occur in spin-transfer-driven magnetization dynamics as well. However, the system is far from equilibrium and the classical notion of spin-waves fails. Indeed, it is the large-amplitude magnetization precession induced by spin transfer that plays the role of reference state, and spin-waves only exist in a generalized, non-equilibrium sense, as small-amplitude perturbations of that state \cite{Bertotti2001a,Kashuba2006,Garanin2009}. This scenario emerges with clarity in the time-dependent vector basis in which the reference magnetization precession is stationary. The spin-wave equations in this basis are characterized by two features: (i) a well-defined dispersion relation $\omega \left ( q; \cos \theta_0 \right )$, whose non-equilibrium nature is revealed by its explicit dependence on the magnetization precession amplitude $\cos \theta_0$; and (ii) the presence of time-periodic coupling terms due to the magnetostatic fields generated by individual spin-waves. This coupling leads to the appearance of narrow instability tongues around the parametric resonance condition $\omega(q; \cos \theta_0) \sim \omega_0$, where $\omega_0$ is the magnetization precession angular frequency.

Spin-wave instabilities occur only for particular combinations of external magnetic field and injected spin-polarized current. In addition, instabilities result in limited spatial and temporal distortions which somewhat obscure but yet do not completely disrupt the precessional character of the original state. This robustness of excited precessions with respect to spin-wave instabilities has a precise physical origin. Indeed, the discrete nature of the spin-wave spectrum caused by boundary conditions in sub-micrometer devices reduces the number of available spin-wave modes which can contribute to instabilities. On the other hand, the strength of the magnetostatic effects responsible for instabilities is drastically reduced, due to the ultra-thin nature of spin-transfer devices, and instability thresholds are consequently enhanced. Finally, spin-transfer-driven precessions are characterized by large amplitudes and, as such, are less easily masked by the onset of non-uniform modes. In spin-transfer nano-oscillators, spin-wave instabilities are expected to result in increased oscillator line-widths, a conclusion that might explain some of the puzzling experimental results obtained in this area \cite{Mistral2006}.

\begin{figure}
\begin{center}
\includegraphics[width=0.42\textwidth]{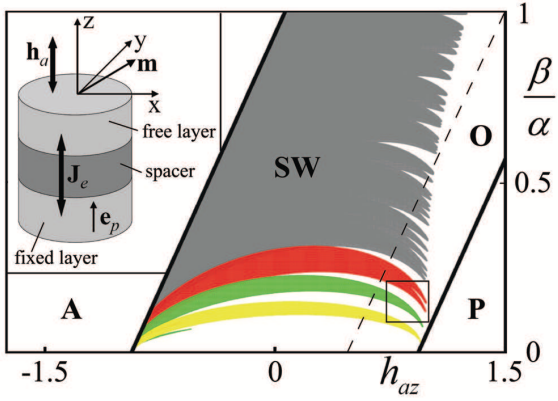}
\end{center}
\caption{\label{FIG: stability 270} (Color online) Stability diagram in $(\haz, \beta/\alpha)$ control plane for a ultra-thin permalloy disk. System parameters: $\alpha = 0.02$, $d = 0.6$, $R = 23.6$, $\Nperp = 0.02$ (lengths are measured in units of the exchange length $\lEX = 5.72$ nm). Magnetization is parallel to spin-polarization in region \textbf{P}; anti-parallel to spin-polarization in region \textbf{A}; precessing around the spin-polarization axis in regions \textbf{O} and \textbf{SW}. Dashed line is an example of line of constant precession amplitude ($\cos \theta_0 = 0.5$) computed from Eq.\eqref{EQ: theta0 beta}. Spin-wave instabilities occur in region \textbf{SW}. Small framed area is shown in detail in Fig. \ref{FIG: numerical 270}. Inset: typical geometry of a nanopillar spin-transfer device.}
\end{figure}

To start the technical discussion, consider a ultra-thin disk with negligible crystal anisotropy (e.g., permalloy). Typically, this disk will be the so-called free layer of a nanopillar spin-transfer device (see inset in Fig. \ref{FIG: stability 270}). The disk plane is parallel to the $(x,y)$ plane and is traversed by a flow of electrons with spin polarization along the $\evec_z$ direction. The dimensionless equation for the dynamics of the normalized magnetization $\mvec (\rvec,t)$ ($|\mvec|^2 = 1$) in the disk in the presence of spin transfer is \cite{Slonczewski1996,Bertotti2005}:

\bea
& & \frac{\partial \mvec}{\partial t} - \alpha \, \mvec \times \frac{\partial \mvec}{\partial t} =
\label{EQ: LLG st} \\
& & - \mvec \times \left ( \haz \evecz + \hvecM + \nabla^2 \mvec - \beta \, \mvec \times \evecz \right ) \,\,\, .
\nonumber
\eea

\noindent Here, the external magnetic field $\haz \evecz$ and the magnetostatic field $\hvecM$ are measured in units of the spontaneous magnetization $\Msat$, time in units of $(\gamma \Msat)^{-1}$ ($\gamma$ is the absolute value of the gyromagnetic ratio), and lengths in units of the exchange length. The external field is perpendicular to the disk plane, while the spin-transfer torque is simply proportional to the sine of the angle between $\mvec$ and $\evecz$. The parameter $\beta$ is proportional to the spin-polarized current density (see \cite{Bertotti2005} for the detailed definition), and in typical situations it is comparable with the damping constant $\alpha$.

Whenever $\left | \haz - \beta / \alpha \right | \le \Npar - \Nperp$ ($\Npar$ and $\Nperp$ are the disk demagnetizing factors, with $\Npar + 2 \Nperp = 1$), Eq.\eqref{EQ: LLG st} admits time-harmonic solutions $\mveco (t)$, corresponding to spatially uniform precession of the magnetization around the $z$-axis \cite{Bazaliy2004} (see Fig. \ref{FIG: stability 270}). The precession amplitude and angular frequency are respectively equal to:

\beq
 \cos \theta_0 = \frac{\haz - \beta/\alpha}{\Npar - \Nperp} \, \, \, , \, \, \, \omega_0 = \frac{\beta}{\alpha} \, \, \, .
 \label{EQ: theta0 beta}
 \eeq

To study the stability of  $\mveco(t)$, consider the perturbed motion $\mvec ( \rvec, t ) = \mveco(t) + \delta \mvec ( \rvec, t )$, with $| \delta \mvec (\rvec, t ) | \ll 1$. The corresponding magnetostatic field will be: $\hvecM ( \rvec, t ) = - \Npar \mvecopar - \Nperp \mvecoperp + \delta \hvecM (\rvec, t )$, where $\delta \hvecM$ represents the magnetostatic field generated by  $\delta \mvec$. Since we are interested in ultra-thin layers, we shall assume that  $\delta \mvec$ does not depend on $z$: $\delta \mvec(\rvec,t) =  \delta \mvec(x,y,t)$.

The perturbation $\delta \mvec$ is orthogonal to $\mveco (t)$ at all times, since the local magnetization magnitude $|\mvec|^2 = 1$ must be preserved. Hence, it is natural to represent $\delta \mvec$ in the time-dependent vector basis $\left ( \evecone(t), \evectwo(t) \right )$ defined in the plane perpendicular to $\mveco(t)$, with $\evectwo (t)$ parallel to $\evecz \times \mveco(t)$ and $\evecone (t)$ such that $\left ( \evecone, \evectwo, \mveco \right )$ form a right-handed orthonormal basis. The perturbation can be written as: $\delta \mvec (\rvec, t ) = \delta m_{1} (\rvec, t ) \, \evecone ( t ) + \delta m_{2} (\rvec, t ) \, \evectwo ( t )$. By linearizing Eq.\eqref{EQ: LLG st} around $\mveco(t)$ and averaging the linearized equation over the layer thickness, one obtains the following coupled differential equations in matrix form:

\bea
&& \left ( \begin{array}{cc} 1 & \alpha \\ - \alpha & 1 \end{array} \right ) \frac{\partial}{\partial t} \left( \begin{array}{c} \delta m_{1} \\ \delta m_{2} \end{array} \right ) = \left ( \begin{array}{cc} 0 & 1 \\ -1 & 0 \end{array} \right ) \left ( \begin{array}{c} \left < \delta \hM \right >_1 \\ \left < \delta \hM \right >_2 \end{array} \right ) +
\nonumber \\
& + & \left ( \begin{array}{cc} 0 & \Nperp + \nabla^{2}_{\bot} \\ - \Nperp - \nabla^{2}_{\bot} & 0 \end{array} \right ) \left ( \begin{array}{c} \delta m_{1} \\ \delta m_{2} \end{array} \right )  \, \, \, ,
\label{EQ: ddm/dt matr}
\eea

\noindent where $\nabla^{2}_\bot = \partial^2 / \partial x^2 + \partial^2 / \partial y^2$, while $\left < \dots \right >$ represents the $z$ average over the thickness of the disk, and $\left < \delta \hM \right >_1 = \left < \delta \hvecM \right > \cdot \evecone ( t )$, $\left < \delta \hM \right >_2 = \left < \delta \hvecM \right > \cdot \evectwo ( t )$.

To grasp the physical consequences of Eq.\eqref{EQ: ddm/dt matr}, consider the plane-wave perturbation $\delta \mvec (\rvec, t) = \avec (t) \exp (i \, \qvec \cdot \rvec )$ in an infinite layer ($\Nperp = 0$). The corresponding magnetostatic field is \cite{Bertotti2009}:

\beq
\left < \delta \hvecM \right > = - s_q \, \delta \mvec_z - \left ( 1 - s_{q} \right ) \delta \mvec_q \, \, \, ; \, \, \,   s_{q} = \frac{1 - \exp (-q d)}{q d} \, \, \, ,
\label{EQ: dhm PW}
\eeq

\noindent where $\delta \mvec_z = \left ( \delta \mvec \cdot \evecz \right ) \evec_z$ and $\delta \mvec_q = \left ( \delta \mvec \cdot \evec_q \right ) \evec_q$, $\evec_q$ being the unit vector in the $\qvec$ direction. The field $- s_q \, \delta \mvec_z$ is generated by the magnetic charges at the layer surface, whereas the field $- \left ( 1 - s_{q} \right ) \delta \mvec_q$ is due to volume charges. By taking into account that $\nabla_{\bot}^2 \delta \mvec = - q^2 \delta \mvec$, $\delta \mvec_z \cdot \evecone (t) = \delta m_1 \sin^2 \theta_0$, and $\delta \mvec_z \cdot \evectwo (t) = 0$, one finds from Eq.\eqref{EQ: ddm/dt matr} that $\mveco(t)$ is always stable with respect to the action of exchange forces and surface magnetic charges. Only volume charges can make the precession unstable. This conclusion follows from the fact that the $z$-axis, along which the surface-charge magnetostatic field is directed, is a symmetry axis for the problem. Surface-charge-driven instabilities may appear in non-uniaxial systems.

The two-dimensional and uniaxial character of the problem makes it natural to introduce polar coordinates $(r, \phi)$ in the disk plane, with the origin at the centre of the disk. The natural boundary condition in polar coordinates is $\left . \partial \delta \mvec / \partial r \right |_{r = R} = 0$, where $R$ is the disk radius. The generic perturbation satisfying this boundary condition consists of cylindrical spin-waves of the type:

\beq
\delta \mvec (r, \phi, t) = \sum_{n = -\infty}^{+\infty} \sum_{k = 0}^{\infty} \avec_{n k} ( t ) \, J_n \left ( q_{n k} r \right ) \, \exp (i n \phi) \, \, \, ,
\label{EQ: dm_series}
\eeq

\noindent where $J_n (z)$ is the $n$-th order Bessel function. The wave-vector amplitude $q_{n k}$ is identified by two subscripts because, for each $n$, it must satisfy the boundary condition $\partial J_n (z) / \partial z = 0$ for $z = q_{n k} R$, which has infinite solutions $q_{n 0} \, , \, q_{n1} \, , \, q_{n2} \, , \dots$ of increasing amplitude. The cylindrical spin-waves $F_{n k} (r, \phi) = J_n \left ( q_{n k} r \right ) \, \exp (i n \phi)$ are a complete orthogonal set of eigenfunctions of the $\nabla^2_{\bot}$ operator: $ \nabla^2_{\bot} \, F_{n k} (r, \phi) = - q_{n k}^2 \, F_{n k} (r, \phi)$. The magnetostatic field $\delta \hvecM$ can be computed by applying Eq.\eqref{EQ: dhm PW} to the plane-wave integral representation: $F_{n k} ( r, \phi ) = 1/( 2 \pi i^n ) \, \int_{0}^{2 \pi} \exp \left ( i \, \qvec_{n k} \cdot \rvec \right ) \, \exp (i n \psi) \, d \psi$, where the polar representation of $\rvec$ and $\qvec_{n k}$ is $\rvec = (r, \phi)$ and $\qvec_{n k} = \left ( q_{n k}, \psi \right ) $, respectively. By following these steps, writing $\avec_{n k} (t)$ as $\avec_{n k} (t) = c_{n k,1} (t) \, \evecone(t) + c_{n k,2} (t) \, \evectwo(t)$, and neglecting small terms proportional to $\Nperp$, Eq.\eqref{EQ: ddm/dt matr} is transformed into the following system of coupled equations:

\bea
& & \frac{d c_{n k}}{d t} = A_{n k} \, c_{n k} + \sum_{p = 0}^{\infty} \, \frac{\Delta_{n k; p}^{+}}{\Delta_{n k}} \, \mathcal{R}_{n+2,p}(t) \, c_{n+2,p}
\label{EQ: dcnk/dt} \\
& + & \sum_{p = 0}^{\infty} \, \frac{\Delta_{n k; p}^{-}}{\Delta_{n k}} \, \mathcal{R}_{n-2,p}^{*}(t) \, c_{n-2,p} \, \, \, ; \, \, \, c_{n k} \equiv  \left ( \begin{array}{c} c_{nk,1} \\ c_{nk,2} \end{array} \right ) \, \, \, ,
\nonumber
\eea

\noindent where:

\beq
A_{n k} = \frac{1}{1 + \alpha^{2}} \left ( \begin{array}{cc} 1 & - \alpha \\
\alpha & 1 \end{array} \right) \left( \begin{array}{cc} 0 & - \nu_{n k} \\ \nu_{n k} - \kappa_{n k} \sin^{2}
\theta_{0} & 0 \end{array} \right ) \, ,
\label{EQ: Ank}
\eeq

\bea
\mathcal{R}_{n k} (t) & = &  \exp \left ( 2 i \omega_0 t \right ) \, \frac{1 - s_{n k}}{4} \, \times
\label{EQ: Rnk} \\
& \times & \frac{1}{1 + \alpha^{2}} \, \left ( \begin{array}{cc} 1 & - \alpha \\
\alpha & 1 \end{array} \right ) \left ( \begin{array}{cc} i \, \cos \theta_{0} & -1 \\ - \cos^{2} \theta_{0}
 & - i \, \cos \theta_{0} \end{array} \right ) \, \, \, ,
\nonumber
\eea

\beq
\Delta_{n k; p}^{\pm} = \int_{0}^{R} r \, J_{n}\left ( q_{n k} r \right ) \, J_{n}\left ( q_{n \pm 2, p} r \right ) \, d r \, \, \, ,
\label{EQ: Delta_nkp}
\eeq

\noindent and $\Delta_{n k} = \int_{0}^{R} r \, J_{n}^2 \left ( q_{n k} r \right ) \, d r$, $\nu_{n k} =  q_{n k}^{2} + \left ( 1 - s_{n k} \right )/2$, $\kappa_{n k} = - 1 + 3 \left ( 1- s_{n k} \right )/2$, $s_{n k}$ being the value of $s_q$ in Eq.\eqref{EQ: dhm PW} for $q = q_{n k}$.

The coupling terms proportional to $\mathcal{R}_{n k} (t)$ in Eq.\eqref{EQ: dcnk/dt} are the consequence of  volume-charge magnetostatic effects. They are all of the order of $\left ( 1 - s_q \right )$. One has that $\left ( 1 - s_q \right ) \ll 1$ up to $q \sim 1$ in ultra-thin layers with $d \lesssim 1$ (see Eq.\eqref{EQ: dhm PW}). If one neglects these terms altogether, one obtains a system of fully decoupled equations for individual cylindrical spin-waves, characterized by the dispersion relation:

\bea
\omega^2 \left ( q; \cos \theta_0 \right ) & = &  \left ( q^2 + \frac{1 - s_q}{2} \right ) \times
\label{EQ: dispersion} \\
& \times & \left ( q^2 + s_q + \frac{1 - 3 s_q}{2} \, \cos^2 \theta_0 \right ) \, \, \, ,
\nonumber
\eea

\noindent which is obtained from Eq.\eqref{EQ: Ank} in the limit $\alpha \rightarrow 0$. However, the time-periodic coupling terms may give rise to parametric instabilities. Interestingly, these instabilities are governed by a small number of dominant terms, which can be identified by using the asymptotic formula $J_n (z) \sim \sqrt{2/\pi z} \, \cos \left ( z - n \pi / 2 - \pi / 4 \right )$ in the equation expressing boundary conditions. One obtains the estimate $q_{n k} \simeq \pi \, (2 s + 1) / 4 R$, where $s = |n| + 2 k$. When this approximate expression is used for $q_{n \pm 2, p}$ in Eq.\eqref{EQ: Delta_nkp}, one obtains:

\bea
n \ge 2 \, & : & \, \, \, \Delta_{n k; p}^{\pm} \simeq \Delta_{n k} \, \delta_{p, k \mp 1} \, \, \, , \, \, \, \Delta_{n 0; p}^{+} \simeq 0 \, \, \, , \nonumber \\
n = \pm 1 \, & : & \, \, \, \Delta_{1k; p}^{-} = \Delta_{-1,k; p}^{+} = \Delta_{1 k} \delta_{p k} \, \, \, ,
\label{EQ: Delta rel} \\
n \le -2 \, & : & \, \, \, \Delta_{n k; p}^{\pm} \simeq \Delta_{n k} \, \delta_{p, k \pm 1} \, \, \, , \, \, \, \Delta_{n 0; p}^{-} \simeq 0 \, \, \, .
\nonumber
\eea

These relations have an important physical consequence, which is best appreciated by rewriting Eq.\eqref{EQ: dm_series} in the form: $\delta \mvec = \sum_{s=0}^{\infty} \delta \mvec^{(s)}$, where:

\beq
\delta \mvec^{(s)} = \displaystyle \sum_{|n| + 2k = s} \avec_{n k} ( t ) \, J_n \left ( q_{n k} r \right ) \, \exp (i n \phi) \, \, \, .
\label{EQ: dm_chain}
\eeq

\noindent Under the approximation \eqref{EQ: Delta rel}, one finds from Eq.\eqref{EQ: dcnk/dt} that $\delta \mvec^{(s_1)}$ is decoupled from $\delta \mvec^{(s_2)}$ for any $s_2 \neq s_1$. On the other hand, for each $s$, the $(s+1)$ cylindrical waves (namely, $n = s, s-2, \dots, -s +2, -s$) involved in $\delta \mvec^{(s)}$ form a one-dimensional chain, in the sense that only neighboring waves in the above list are coupled. The absence of coupling between distinct chains would be complete if the approximation $q_{n k} \simeq \pi \, (2 s + 1) / 4 R$ were exact. In that case, all the cylindrical waves in $\delta \mvec^{(s)}$ would be characterized by exactly the same wave-vector amplitude.

\begin{figure}
\begin{center}
\includegraphics[width=0.45\textwidth]{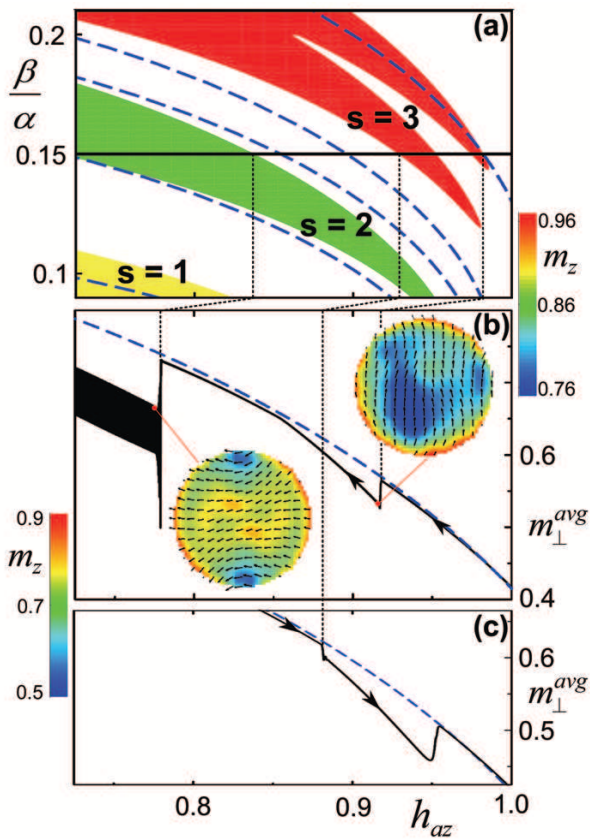}
\end{center}
\caption{\label{FIG: numerical 270} (Color online) (a): Magnification of Fig. \ref{FIG: stability 270}. Labels $s = 1,2,3$ identify the perturbation chain responsible for the corresponding instability tongue. The pair of dashed lines accompanying each of the $s = 2$ and $s = 3$ tongues (one line only for $s = 1$) represents the parametric resonance condition $\omega\left (q_{n k}; \cos \theta_0 \right ) = \omega_0$ for the largest and smallest $q_{n k}$ in the chain. Horizontal line at $\beta / \alpha = 0.15$ is line along which the computer simulations shown in (b) and (c) were carried out. (b) and (c): Magnitude $m_{\bot}^{\text{avg}}$ of average in-plane magnetization obtained from numerical integration of Eq.\eqref{EQ: LLG st} under decreasing (b) and increasing (c) external magnetic field. Dashed line represents the prediction of Eq.\eqref{EQ: theta0 beta} for $\sin \theta_0$. Snapshots illustrate magnetization patterns appearing just after the instability jumps. Vertical dotted lines are guides for the eye to compare thresholds with theoretical predictions obtained from (a).}
\end{figure}

Instabilities are governed by the multipliers of the one-period map \cite{Perko1996} associated with the dynamics of $\delta \mvec^{(s)}$. We have used Eqs.\eqref{EQ: dcnk/dt} and \eqref{EQ: Delta rel} to make a numerical study of these multipliers for different chains, in order to obtain the instability pattern associated with each of them. The results for a permalloy disk with radius $R = 135$ nm and thickness $d = 3.43$ nm are shown in Fig. \ref{FIG: stability 270}. The band (\textbf{O} + \textbf{SW}) in between the \textbf{P} and \textbf{A} regions is where magnetization precession occurs. Spin-wave instabilities appear in region \textbf{SW}. Each chain $\delta \mvec^{(s)}$ provides a distinct instability channel. In particular, chains $s = 1$, $s = 2$, and $s = 3$ ($s = 0$ yields no instability at all) give rise to well-separated instability regions that can be neatly resolved, as shown in Fig. \ref{FIG: numerical 270}(a). In general, the $s$-th chain gives rise to an instability tongue around the parametric resonance condition $\omega \left ( q; \cos \theta_0 \right ) \sim \omega_0$, where $\omega \left ( q; \cos \theta_0 \right )$ is given by Eq.\eqref{EQ: dispersion} and $q$ is of the order of the wave-vector amplitudes involved in the chain (see dashed lines in Fig. \ref{FIG: numerical 270}(a)). According to parametric resonance theory, resonance occurs for $\omega = n \, \omega_0 / 2$, $n = 1, 2, \dots$. The dominant, lowest-threshold resonance occurs for $n = 1$, that is, at $\omega = \omega_0 / 2$. However, one can see from Eq.\eqref{EQ: Rnk} that the parametric frequency is $2 \omega_0$ rather than $\omega_0$, which explains why the resonance condition is $\omega \sim \omega_0$. This peculiarity is the consequence of the rotational invariance of the problem, and is expected to disappear in situations with broken rotational symmetry. Interestingly, Fig. \ref{FIG: stability 270} reveals that, for a given precession amplitude $\cos \theta_0$ (see dashed line), applying larger fields and currents has a stabilizing effect on the precession. Also, larger fields stabilize precessions of given frequency $\omega_0 = \beta / \alpha$.

To test the predictions of the theory, we have carried out computer simulations based on the numerical integration of Eq.\eqref{EQ: LLG st} by the methods discussed in Ref. \cite{DAquino2005a}. Simulations were carried out by slowly varying the external magnetic field under constant current. As shown in Fig. \ref{FIG: numerical 270}(b), at large fields the magnitude $m_{\bot}^{\text{avg}}$ of the average in-plane magnetization is in full agreement with the prediction of Eq.\eqref{EQ: theta0 beta} for spatially uniform precession. Then, under decreasing field $m_{\bot}^{\text{avg}}$ exhibits well-pronounced jumps, whose positions agree with the theoretical instability thresholds for the $s=2$ and $s=3$ chains within ten percent. Beyond these jumps, non-uniform modes appear in the dynamics (see Fig. \ref{FIG: numerical 270}(b)), characterized by two-fold and three-fold patterns that are consistent with the symmetry of the cylindrical waves involved in the $s = 2$ and $s = 3$ chains, respectively. Agreement with the theory is also confirmed by the hysteresis in the instability thresholds occurring under decreasing or increasing external field (Fig. \ref{FIG: numerical 270}(c)).

The stability of spin-transfer-driven magnetization precessions has been studied in this Letter under the simplest conditions, namely, uniaxial symmetry and pure $\sin \theta_0$ angular dependence of the spin-torque. Several features of physical interest have emerged: the role played by non-equilibrium spin-waves; the fact that instabilities are governed by distinct chains of magnetostatically coupled spin-waves; and the fact that excited precessions are not completely disrupted but only somewhat obscured by spin-wave instabilities. Future work will be devoted to extending the present approach to more general, non-uniaxial geometries.


\begin{thebibliography}{99}
\expandafter\ifx\csname natexlab\endcsname\relax\def\natexlab#1{#1}\fi
\expandafter\ifx\csname bibnamefont\endcsname\relax
  \def\bibnamefont#1{#1}\fi
\expandafter\ifx\csname bibfnamefont\endcsname\relax
  \def\bibfnamefont#1{#1}\fi
\expandafter\ifx\csname citenamefont\endcsname\relax
  \def\citenamefont#1{#1}\fi
\expandafter\ifx\csname url\endcsname\relax
  \def\url#1{\texttt{#1}}\fi
\expandafter\ifx\csname urlprefix\endcsname\relax\def\urlprefix{URL }\fi
\providecommand{\bibinfo}[2]{#2}
\providecommand{\eprint}[2][]{\url{#2}}

\bibitem[{\citenamefont{Slonczewski}(1996)}]{Slonczewski1996}
\bibinfo{author}{\bibfnamefont{J.~C.} \bibnamefont{Slonczewski}},
  \bibinfo{journal}{J. Magn. Magn. Mater.} \textbf{\bibinfo{volume}{159}},
  \bibinfo{pages}{L1} (\bibinfo{year}{1996}).

\bibitem[{\citenamefont{Berger}(1996)}]{Berger1996}
\bibinfo{author}{\bibfnamefont{L.}~\bibnamefont{Berger}},
  \bibinfo{journal}{Phys. Rev. B} \textbf{\bibinfo{volume}{54}},
  \bibinfo{pages}{9353} (\bibinfo{year}{1996}).

\bibitem[{\citenamefont{Kiselev et~al.}(2003)\citenamefont{Kiselev, Sankey,
  Krivorotov, Emley, Schoelkopf, Buhrman, and Ralph}}]{Kiselev2003}
\bibinfo{author}{\bibfnamefont{S.~I.} \bibnamefont{Kiselev}}
  \bibnamefont{et} \bibinfo{author}{\bibfnamefont{al.}},
  \bibinfo{journal}{Nature} \textbf{\bibinfo{volume}{425}},
  \bibinfo{pages}{380} (\bibinfo{year}{2003}).

\bibitem[{\citenamefont{Rippard et~al.}(2004)\citenamefont{Rippard, Pufall,
  Kaka, Russek, and Silva}}]{Rippard2004}
\bibinfo{author}{\bibfnamefont{W.~H.} \bibnamefont{Rippard}},
  \bibinfo{author}{\bibfnamefont{M.~R.} \bibnamefont{Pufall}},
  \bibinfo{author}{\bibfnamefont{S.}~\bibnamefont{Kaka}},
  \bibinfo{author}{\bibfnamefont{S.~E.} \bibnamefont{Russek}},
  \bibnamefont{and} \bibinfo{author}{\bibfnamefont{T.~J.} \bibnamefont{Silva}},
  \bibinfo{journal}{Phys. Rev. Lett.} \textbf{\bibinfo{volume}{92}},
  \bibinfo{pages}{027201} (\bibinfo{year}{2004}).

\bibitem[{\citenamefont{Krivorotov et~al.}(2007)\citenamefont{Krivorotov,
  Berkov, Gorn, Emley, Sankey, Ralph, and Buhrman}}]{Krivorotov2007}
\bibinfo{author}{\bibfnamefont{I.~N.} \bibnamefont{Krivorotov}}
  \bibnamefont{et} \bibinfo{author}{\bibfnamefont{al.}},
  \bibinfo{journal}{Phys. Rev. B} \textbf{\bibinfo{volume}{76}},
  \bibinfo{pages}{024418} (\bibinfo{year}{2007}).

\bibitem[{\citenamefont{Boone et~al.}(2009)\citenamefont{Boone, Katine,
  Childress, Tiberkevich, Slavin, Zhu, Cheng, and Krivorotov}}]{Boone2009}
\bibinfo{author}{\bibfnamefont{C.~T.} \bibnamefont{Boone}}
  \bibnamefont{et} \bibinfo{author}{\bibfnamefont{al.}},
  \bibinfo{journal}{Phys. Rev. Lett.} \textbf{\bibinfo{volume}{103}},
  \bibinfo{pages}{167601} (\bibinfo{year}{2009}).

\bibitem[{\citenamefont{Suhl}(1957)}]{Suhl1957}
\bibinfo{author}{\bibfnamefont{H.}~\bibnamefont{Suhl}}, \bibinfo{journal}{J.
  Phys. Chem. Solids} \textbf{\bibinfo{volume}{1}}, \bibinfo{pages}{209}
  (\bibinfo{year}{1957}).

\bibitem[{\citenamefont{Bertotti et~al.}(2001)\citenamefont{Bertotti,
  Mayergoyz, and Serpico}}]{Bertotti2001a}
\bibinfo{author}{\bibfnamefont{G.}~\bibnamefont{Bertotti}},
  \bibinfo{author}{\bibfnamefont{I.~D.} \bibnamefont{Mayergoyz}},
  \bibnamefont{and} \bibinfo{author}{\bibfnamefont{C.}~\bibnamefont{Serpico}},
  \bibinfo{journal}{Phys. Rev. Lett.} \textbf{\bibinfo{volume}{87}},
  \bibinfo{pages}{217203} (\bibinfo{year}{2001}).

\bibitem[{\citenamefont{Kashuba}(2006)}]{Kashuba2006}
\bibinfo{author}{\bibfnamefont{A.}~\bibnamefont{Kashuba}},
  \bibinfo{journal}{Phys. Rev. Lett.} \textbf{\bibinfo{volume}{96}},
  \bibinfo{pages}{047601} (\bibinfo{year}{2006}).

\bibitem[{\citenamefont{Garanin and Kachkachi}(2009)}]{Garanin2009}
\bibinfo{author}{\bibfnamefont{D.~A.} \bibnamefont{Garanin}} \bibnamefont{and}
  \bibinfo{author}{\bibfnamefont{H.}~\bibnamefont{Kachkachi}},
  \bibinfo{journal}{Phys. Rev. B} \textbf{\bibinfo{volume}{80}},
  \bibinfo{pages}{014420} (\bibinfo{year}{2009}).

\bibitem[{\citenamefont{Mistral et~al.}(2006)\citenamefont{Mistral, Kim,
  Devolder, Crozat, Chappert, Katine, Carey, and Ito}}]{Mistral2006}
\bibinfo{author}{\bibfnamefont{Q.}~\bibnamefont{Mistral}},
  \bibnamefont{et} \bibinfo{author}{\bibfnamefont{al.}},
  \bibinfo{journal}{Appl. Phys. Lett.} \textbf{\bibinfo{volume}{88}},
  \bibinfo{pages}{192507} (\bibinfo{year}{2006}).

\bibitem[{\citenamefont{Bertotti et~al.}(2005)\citenamefont{Bertotti, Serpico,
  Mayergoyz, Magni, d'Aquino, and Bonin}}]{Bertotti2005}
\bibinfo{author}{\bibfnamefont{G.}~\bibnamefont{Bertotti}}
  \bibnamefont{et} \bibinfo{author}{\bibfnamefont{al.}},
  \bibinfo{journal}{Phys. Rev. Lett.} \textbf{\bibinfo{volume}{94}},
  \bibinfo{pages}{127206} (\bibinfo{year}{2005}).

\bibitem[{\citenamefont{Bazaliy et~al.}(2004)\citenamefont{Bazaliy, Jones, and
  Zhang}}]{Bazaliy2004}
\bibinfo{author}{\bibfnamefont{Y.~B.} \bibnamefont{Bazaliy}},
  \bibinfo{author}{\bibfnamefont{B.~A.} \bibnamefont{Jones}}, \bibnamefont{and}
  \bibinfo{author}{\bibfnamefont{S.~C.} \bibnamefont{Zhang}},
  \bibinfo{journal}{Phys. Rev. B} \textbf{\bibinfo{volume}{69}},
  \bibinfo{pages}{094421} (\bibinfo{year}{2004}).

\bibitem[{\citenamefont{Bertotti et~al.}(2009)\citenamefont{Bertotti,
  Mayergoyz, and Serpico}}]{Bertotti2009}
\bibinfo{author}{\bibfnamefont{G.}~\bibnamefont{Bertotti}},
  \bibinfo{author}{\bibfnamefont{I.~D.} \bibnamefont{Mayergoyz}},
  \bibnamefont{and} \bibinfo{author}{\bibfnamefont{C.}~\bibnamefont{Serpico}},
  \emph{\bibinfo{title}{Nonlinear Magnetization Dynamics in Nanosystems}}
  (\bibinfo{publisher}{Elsevier}, \bibinfo{address}{Oxford},
  \bibinfo{year}{2009}),
   \bibinfo{pages}{Sect. 8.5}.

\bibitem[{\citenamefont{Perko}(1996)}]{Perko1996}
\bibinfo{author}{\bibfnamefont{L.}~\bibnamefont{Perko}},
  \emph{\bibinfo{title}{Differential Equations and Dynamical Systems}}
  (\bibinfo{publisher}{Springer}, \bibinfo{address}{New York},
  \bibinfo{year}{1996}).

\bibitem[{\citenamefont{d'Aquino et~al.}(2005)\citenamefont{d'Aquino, Serpico,
  and Miano}}]{DAquino2005a}
\bibinfo{author}{\bibfnamefont{M.}~\bibnamefont{d'Aquino}},
  \bibinfo{author}{\bibfnamefont{C.}~\bibnamefont{Serpico}}, \bibnamefont{and}
  \bibinfo{author}{\bibfnamefont{G.}~\bibnamefont{Miano}}, \bibinfo{journal}{J.
  Comput. Phys.} \textbf{\bibinfo{volume}{209}}, \bibinfo{pages}{730}
  (\bibinfo{year}{2005}).

\end{thebibliography}

 \end{document}